# An Intuitive Model for Illustrating the Classical Doppler Effect Equation

Hangtian Lei

***Abstract*:** The Doppler effect has many applications in science and engineering fields. Although the format of the classical Doppler effect equation is simple, the derivation for the equation in physics textbooks is not intuitive to many students. This article provides a simple but effective model to illustrate the derivation for the classical Doppler effect equation. This model visualizes frequency, wavelength, and wave speed, which helps students to establish an intuitive and clear understanding of mechanical waves. This model also provides a clear visualization that the wave speed is unchanging in an underlying medium, which is a key point in understanding the classical Doppler effect.

## I. INTRODUCTION

The Doppler effect is an important topic in university and high school physics classes. Although the format of the classical Doppler effect equation is simple, the derivation for the equation in current physics textbooks is not intuitive to many students. Students commonly confuse the situation of source moving with the situation of observer moving. Furthermore, many students feel it is not intuitive to understand the situation when the source is moving.

The most salient obstacle to the understanding of the Doppler effect is that we could not visualize waves (e.g., sound waves) in our daily life, except in some limited situations (e.g., water waves). As a result, we do not have a clear visualization of frequency, wavelength, and wave speed. A key point in understanding the classical Doppler effect is that the wave speed only depends on the underlying medium, and is independent from the source or observer's speed. To understand the essence of the classical Doppler effect, it is important to have a clear visualization of wavelength and frequency, as well as an intuitive understanding of the unchanging wave speed principle. Our life experience with water waves could not provide us with such a clear visualization and intuitive understanding.

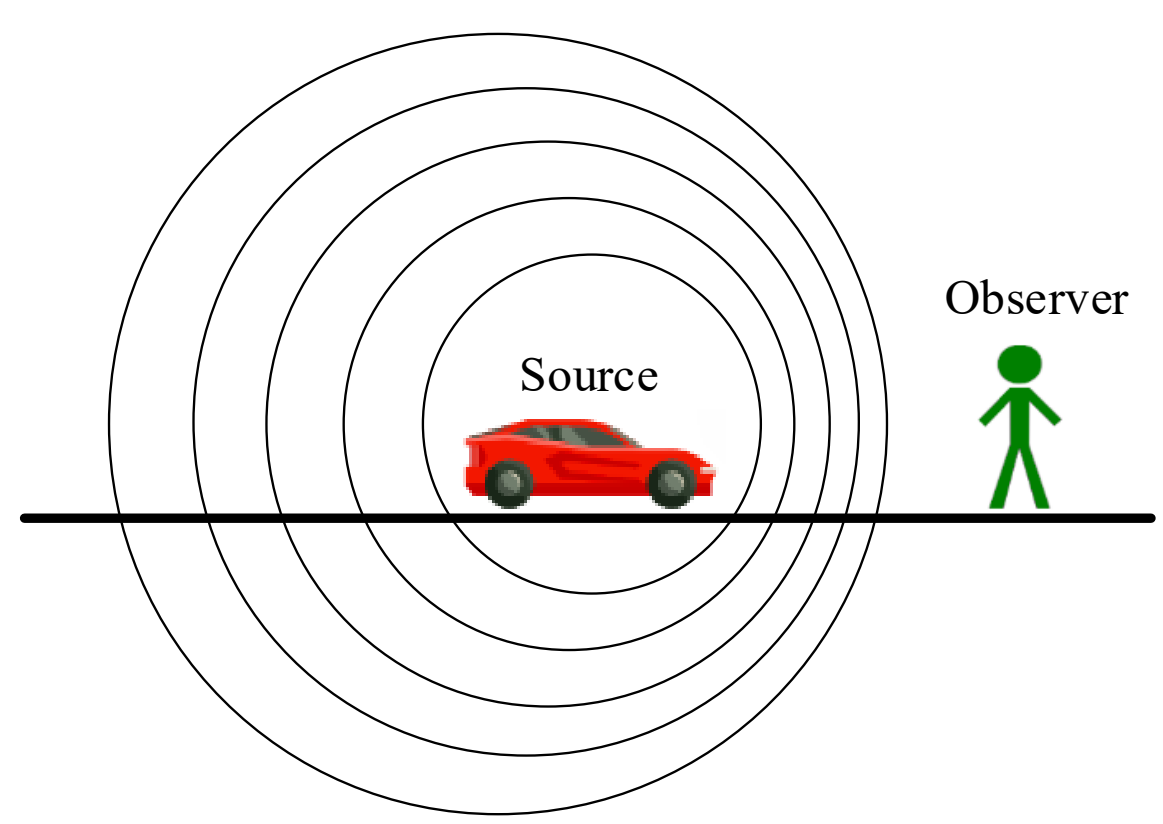

**Fig. 1.** Representation of waves as a group of circles.

Hangtian Lei is an Associate Professor in the Department of Electrical and Computer Engineering at the University of Idaho, Moscow, Idaho, USA, where he has been working since 2017. Email: hlei7@uidaho.edu.

Illustration of the classical Doppler effect in physics textbooks [1, 2] typically uses a group of circles to represent waves, as shown in Fig. 1. Such a representation cannot provide students with an intuitive understanding that the wave speed is unchanging in an underlying medium (e.g., the air). Students are likely to have a misconception that the wave speed will be affected if the source is moving. Some authors use a ball passing or a bullet shooting model [3, 4] to illustrate the classical Doppler effect, and assume that the speed of a ball or a bullet is independent from the source speed. Such assumption is not realistic and not intuitive.

Considering the limitations of the illustration in current textbooks and articles, an intuitive model is proposed in this article for illustrating the classical Doppler effect equation with better clarity.

## II. THE PROPOSED ILLUSTRATION MODEL

The proposed illustration model includes two people standing (or moving) beside a belt, as shown in Fig. 2. One of the two people is the source, the other one is the observer. The belt is powered by a pair of motors operating at a constant speed of $v$. The source marks dots on the belt using a marker pen, at a frequency of $f_s$. Thus, the time between the source marking two consecutive dots is $1/f_s$. *It should be noted that the action of marking dots does not affect the belt speed. In other words, the belt is always moving at a constant speed.*

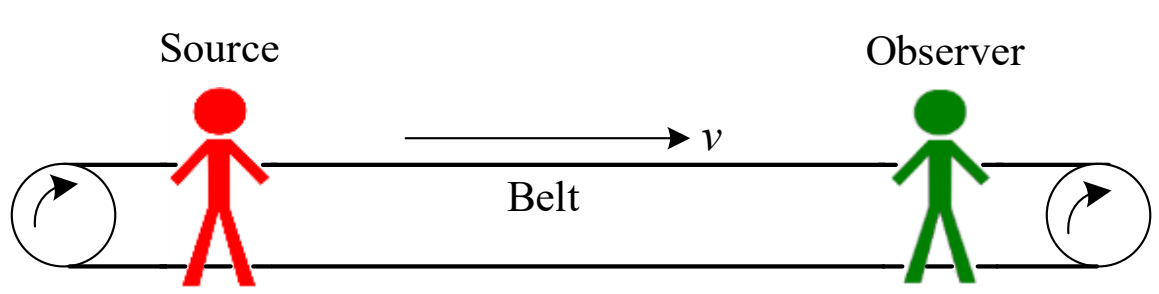

**Fig. 2.** The belt model.

As the belt is moving, a series of dots will be marked on the belt, as shown in Fig. 3. The distance between two consecutive dots is $\lambda$, which represents the wavelength. When a dot passes the observer, we say that the observer receives a dot. The observer receives dots at a frequency of $f_o$, which represents observer frequency of a wave. Analogy between this belt model and an actual mechanical wave (e.g., sound wave) is listed in Table I.

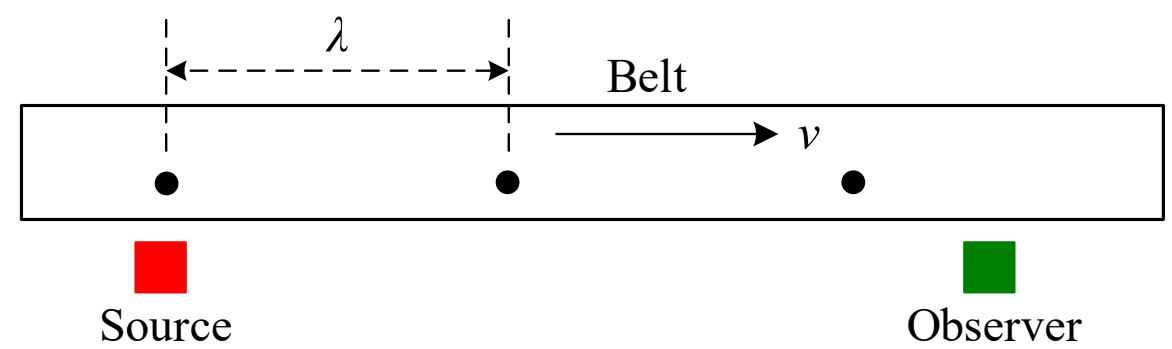

**Fig. 3.** A series of dots marked on the belt.

**TABLE I.** Analogy between the belt model and an actual wave.

| Notation | The Belt Model | Actual Wave |
|---|---|---|
| $f_s$ | Number of dots marked by the source/second | Actual frequency of the wave |
| $f_o$ | Number of dots received by the observer/second | Observer frequency of the wave |
| $v$ | Belt speed | Wave speed |
| $v_s$ | Source speed | Source speed |
| $v_o$ | Observer speed | Observer speed |
| $\lambda$ | Distance between two consecutive dots | Wavelength |

### A. Both source and observer stationary

This scenario is simple, as shown in Fig. 4 (a). Because the time between the source marking two consecutive dots is $1/f_s$, the distance between two consecutive dots,

$$\lambda = \frac{v}{f_s} \tag{1}$$

The observer will receive dots at the same frequency as the source marks the dots. Thus,

$$f_o = f_s \tag{2}$$

For example, the belt moves 12 m/s. The source marks 2 dots/s on the belt. $f_s$ = 2 Hz. The time between the source marking two consecutive dots is 0.5 second. The distance between two consecutive dots is 6 m. The observer receives 2 dots/s. $f_o$ = 2 Hz.

**B. Source moving and observer stationary**

This is the scenario students typically have confusion. With the proposed belt model, the wavelength, frequency, and wave speed can be clearly visualized, which will facilitate students' understanding.

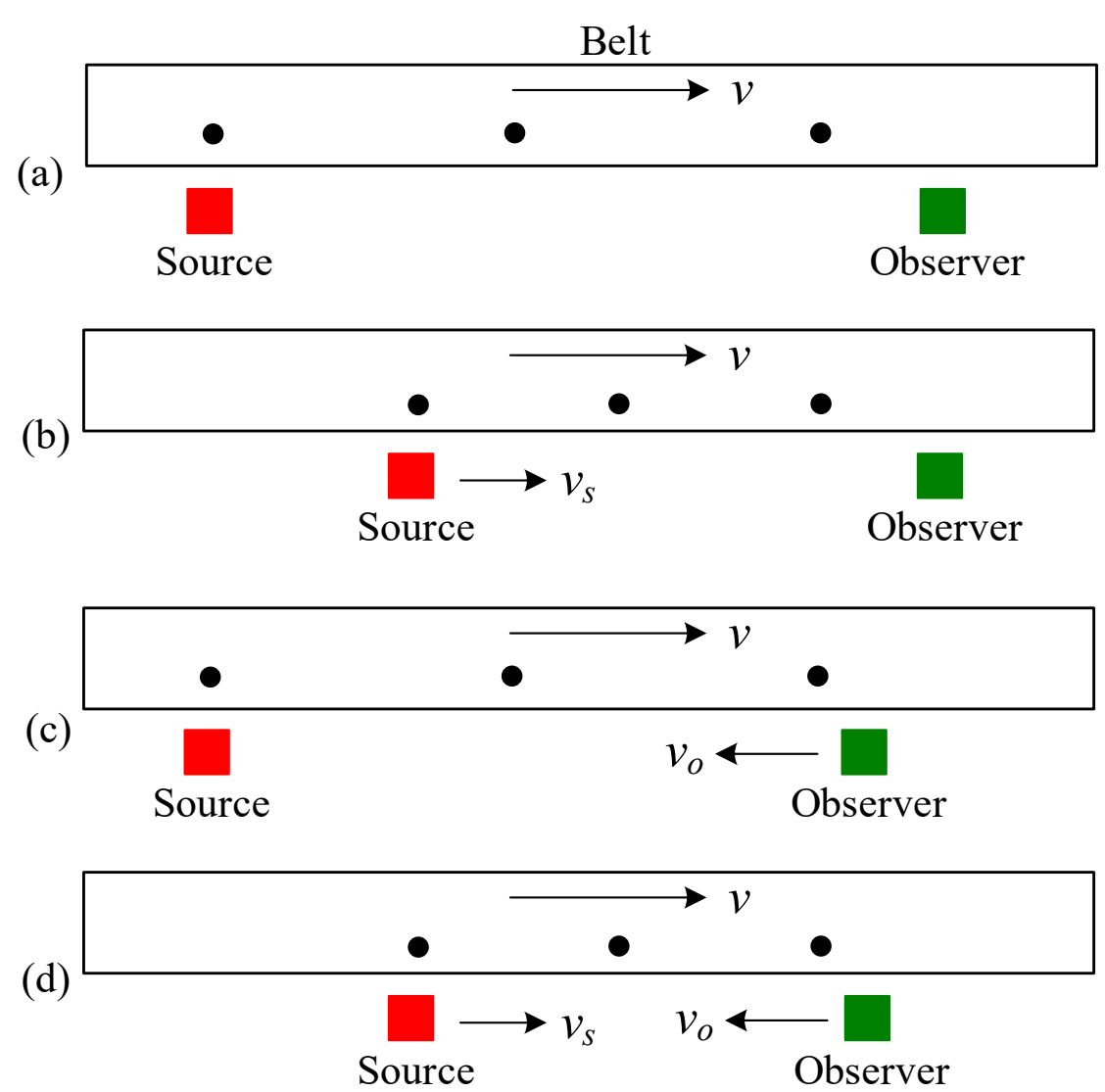

**Fig. 4.** Four typical scenarios. (a) Both source and observer stationary. (b) Source moving and observer stationary. (c) Source stationary and observer moving. (d) Source and observer moving towards each other.

We will illustrate the situation when the source is moving towards the observer. We will only discuss the situation when the source speed is less than the belt speed ($v_s < v$). When the source speed is greater than or equal to the wave speed ($v_s \geq v$), a different phenomenon called "shock wave" will be observed, which is beyond the scope of this article.

Because the source is moving, when a dot is marked on the belt, the distance to its previous dot is shortened compared to the situation when the source is stationary. In other words, compared to Fig. 4 (a), the $\lambda$ in Fig. 4 (b) is shortened due to the source movement. As a result, the observer in Fig. 4 (b) will receive more dots per second compared to Fig. 4 (a), which means the observer frequency of the wave is increased.

In this scenario, the source still marks dots at a frequency of $f_s$. The time between the source marking two consecutive dots is $1/f_s$. Because the source is moving in the same direction as the belt, the distance between two consecutive dots,

$$\lambda = \frac{v - v_s}{f_s} \tag{3}$$

The belt still moves at a constant speed of $v$. Because the observer is stationary, the belt passes the observer at a speed of $v$. Considering the distance between two consecutive dots, the observer receives dots at a frequency of $v/\lambda$. Thus,

$$f_o = \frac{v}{\lambda} = \frac{v f_s}{v - v_s} = \frac{f_s}{1 - v_s/v} \tag{4}$$

For example, the belt moves 12 m/s. The source moves 4 m/s, in the same direction as the belt. The source marks 2 dots/s on the belt. $f_s$ = 2 Hz. The time between the source marking two consecutive dots is 0.5 second. The distance between two consecutive dots is 4 m. The belt passes the observer at a speed of 12 m/s, which means the observer receives 3 dots/s on average. Therefore, $f_o$ = 3 Hz.

If the source is moving away from the observer, the minus sign in Eq. (4) changes to plus.

### C. Source stationary and observer moving

We will illustrate the situation when the observer is moving towards the source. In this scenario, the source still marks dots at a frequency of $f_s$. The time between the source marking two consecutive dots is $1/f_s$. The distance between two consecutive dots,

$$\lambda = \frac{v}{f_s} \tag{5}$$

The belt still moves at a constant speed of $v$. The observer moves in the opposite direction as the belt. The belt passes the observer at a speed of $(v + v_o)$. Considering the distance between two consecutive dots, the observer receives dots at a frequency of $(v + v_o)/\lambda$.

$$f_o = \frac{v+v_o}{\lambda} = \frac{(v+v_o)f_s}{v} = (1 + \frac{v_o}{v})f_s \tag{6}$$

For example, the belt moves 12 m/s. The observer moves 4 m/s towards the source. The source marks 2 dots/s on the belt. $f_s = 2$ Hz. The time between the source marking two consecutive dots is 0.5 second. The distance between two consecutive dots is 6 m. The belt passes the observer at a speed of 16 m/s, which means the observer receives 2.67 dots/s on average. $f_o = 2.67$ Hz.

If the observer is moving away from the source, the plus sign in Eq. (6) changes to minus.

### D. Both source and observer moving

We will illustrate the scenario when the source and observer are moving towards each other. This scenario can be considered as the combination of the two scenarios previously illustrated. With similar procedures illustrated in Sections II.B and II.C, Eq. (7) can be obtained.

$$f_o = f_s \frac{1+v_o/v}{1-v_s/v} = f_s \frac{v + v_o}{v - v_s} \tag{7}$$

For example, the belt moves 12 m/s. The source moves 4 m/s, in the same direction as the belt. The observer moves 4 m/s, in the opposite direction as the belt. The source marks 2 dots/s on the belt. $f_s = 2$ Hz. The time between the source marking two consecutive dots is 0.5 second. The distance between two consecutive dots is 4 m. The belt passes the observer at a speed of 16 m/s, which means the observer receives 4 dots/s on average. $f_o = 4$ Hz.

Eq. (7) is also the general form of classical Doppler effect equation we see in physics textbooks. If the observer moves away from the source, the plus sign in the numerator changes to minus. If the source moves away from the observer, the minus sign in the denominator changes to plus.

## III. APPLICABILITY OF THE MODEL

The belt model proposed in this article can only be used to illustrate the classical Doppler effect of mechanical waves (e.g., sound waves or water waves) whose speed is relative to an underlying medium. For the illustration of relativistic Doppler effect of light and other electromagnetic waves whose speed is relative to the observer instead of an underlying medium, this belt model is not applicable.

## IV. CONCLUSION

A simple and intuitive model is proposed in this article to illustrate the derivation of the classical Doppler effect equation. This model provides a clear visualization of frequency, wavelength, and wave speed, which is very helpful for students to establish a clear understanding of the physics concepts. This model also provides students with an intuitive understanding that the wave speed is unchanging in an underlying medium, which is a key point in understanding the classical Doppler effect.

### V. AUTHOR DECLARATIONS

The author has no conflicts to disclose.

## References

[1] D. Halliday, R. Resnick, and J. Walker, *Fundamentals of Physics*, 10th Ed. (John Wiley & Sons, Hoboken, NJ, 2014), pp. 498–502.

[2] P. G. Hewitt, *Conceptual Physics*, 13th Ed. (Pearson, Essex, UK, 2023), pp. 436.

[3] L. P. S. Kaura and P. Pathak, "A pedagogical model for the Doppler effect with application to sources with constant accelerations," *The Physics Teacher*, vol. 55, pp. 36-37, 2017.

[4] R. Rojas and G. Fuster, "Graphical representation of the Doppler shift: classical and relativistic," *The Physics Teacher*, vol. 45, pp. 306-309, 2007.